# Gate-Defined Quantum Confinement in InSe-based van der Waals Heterostructures

*Matthew Hamer\*[1, 2], Endre Tóvári[2], Mengjian Zhu[1], Michael D. Thompson[3], Alexander Mayorov[4], Jonathon Prance[3], Yongjin Lee[5], Richard P. Haley[3], Zakhar R. Kudrynskyi[6], Amalia Patanè[6], Daniel Terry[1, 2], Zakhar D. Kovalyuk[7], Klaus Ensslin[5], Andrey V. Kretinin[2,8], Andre Geim[1], Roman Gorbachev\*[1, 2],*

[1]School of Physics, University of Manchester, Oxford Road, Manchester, M13 9PL, UK

[2]National Graphene Institute, University of Manchester, Oxford Road, Manchester, M13 9PL, UK

[3]Department of Physics, University of Lancaster, Bailrigg, Lancaster, LA1 4YW, UK

[4]Centre for Advanced 2D Materials, National University of Singapore, 6 Science Drive 2, Singapore

[5]Solid State Physics Laboratory, ETH Zurich, Otto-Stern-Weg 1, 8093 Zürich, Switzerland

[6]School of Physics and Astronomy, University of Nottingham, NG7 2RD, UK

[7]National Academy of Sciences of Ukraine, Institute for Problems of Materials Science, UA-58001, Chernovtsy, Ukraine

Indium selenide, a post-transition metal chalcogenide, is a novel two-dimensional (2D) semiconductor with interesting electronic properties. Its tunable band gap and high electron mobility have already attracted considerable research interest. Here we demonstrate strong quantum confinement and manipulation of single electrons in devices made from few-layer crystals of InSe using electrostatic gating. We report on gate-controlled quantum dots in the Coulomb blockade regime as well as one-dimensional quantization in point contacts, revealing multiple plateaus. The work represents an important milestone in the development of quality devices based on 2D materials and makes InSe a prime candidate for relevant electronic and optoelectronic applications.

The electronic structure of two-dimensional (2D) metal chalcogenides (MCs) depends strongly upon the number of atomic layers. In many MCs the bandgap can vary by as much as 1 eV and its type can change from direct to indirectF10[1,2]. This opens many possibilities for band gap engineering in the construction of complex electronic systems using the van der Waals heterostructure platform[3,4]. In the last six years a large number of devices have been made from few-layer MCs including photodetectors[5], light emitting diodes[6], field effect transistors[7,8] and indirect exciton devices[9], to name but a few. There has also been a great deal of interest in charge confinement within two-dimensional materials including one-dimensional (1D) channels in quantum point contact (QPCs)[10] and zero-dimensional quantum dots (QDs)[11]. Realizing such systems could lead to novel quantum systems including spin-valley qubits[12,13] and Luttinger liquids[14]. The first laterally confined devices realized using 2D crystals were quantum dots fabricated by etching graphene[15,16]. Those dots, however, suffered from limited performance and poor reproducibility due to edge states and charge inhomogeneities introduced by plasma etching[17]. The latter problem can be bypassed in semiconducting 2D crystals where the presence of a band gap enables QDs to be electrostatically defined using gate electrodes. Successful examples of gated defined quantum dots have been reported in 2D transition metal dichalcogenides (TMDCs), i.e. $WSe_2$, $WS_2$ and most recently $MoS_2$ [18–20].

By contrast, 1D quantization has proven elusive: until very recently, graphene has been the only 2D material to display signs of quantized conductance[21–23]. The difficulties in creating QPCs that exhibit quality quantization are due to the following. First, series contact resistances must be minimized to prevent 1D conductance steps from being obscured. Second, charge transport needs to be ballistic with the mean free path exceeding the size of QPC constrictions. Finally, the Fermi wavelength $\lambda_F$ must be comparable to the constriction size, which typically requires low charge densities. Therefore, the constriction size must be very small or the charge carrier mobility $\mu$ very high: an imposing challenge from a fabrication perspective. In practical terms, these requirements rule out many 2D materials. Among the various TMDCs only a handful have shown sufficiently high $\mu$, to observe 1D quantization. The possible contenders are few-layer $WSe_2$ and $WS_2$ which exhibit $\mu$ of around 500 $cm^2$/Vs[24,25] and black phosphorus which has $\mu$ up to ~5,000 cm$^2$/Vs[26] at T=4 K. Recently, however, improvements in device fabrication have led to field-effect transistors made from MoS$_2$ and InSe, which had µ over 20,000 cm$^2$/Vs[27,28], high enough to observe the quantum Hall effect. Such mobilities make these materials promising candidates for the observation of 1D quantization. Indeed, QPC conductance plateaus have been reported in high-quality MoS$_2$ channels[20,29,30]. As for 2D InSe, this metal chalcogenide has a semiconducting bandgap ranging from 1.25 eV in the bulk to 2.9 eV in single-layer samples[28,31]. In addition, the bandgap remains quasi-direct down to few-layer thickness[32], making InSe-based devices promising for coupling with optics[33,34]. Less fortunately, 2D InSe degrades under ambient conditions and, therefore, its exposure to air must be avoided, which requires fabrication in an inert environment[28].

In this report, we describe the first low-dimensional InSe devices defined by local electrostatic gating. To avoid the degradation of the layers under ambient conditions, few-layer InSe crystals were encapsulated in hexagonal boron nitride (hBN) using the dry peel transfer technique inside an argon-filled glovebox chamber[35]. The top hBN encapsulation layer also serves as a dielectric for the top gates deposited later. Ohmic contacts to 2D InSe were formed by depositing the InSe crystals on top of two graphene flakes[27,36], which subsequently were contacted by Cr/Au electrodes as shown in Fig. 1a. To minimize the contact resistance, additional top gates were deposited above each InSe/graphene interface to increase the carrier density and suppress the Schottky barrier. The QDs and QPCs were electrostatically defined with a series of top gates depicted in Fig. 1b. The overall carrier density $n$ was controlled by the global back gate formed by the heavily doped silicon substrate. 2D InSe exhibits highest µ for a thickness of 5-6 layers[28] and, accordingly, we focus below on results obtained from devices with these thicknesses. The onset of Shubnikov-de Haas oscillations observed in devices (see Supplementary Information) yields µ in the order of 10,000 cm$^2$/Vs at T=1.5 K for electron densities of $n \sim 5 \times 10^{12}$ cm$^{-2}$, in agreement with the previous reported values[28].

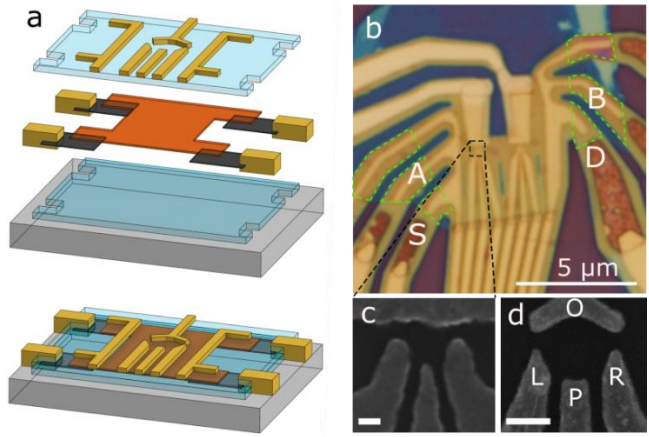

**Figure 1.** hBN/InSe/hBN heterostructure with graphene contacts and multiple top gates. **(a)** Schematic showing individual layers: 2D InSe (red), graphene (dark grey), hBN (blue), Cr/Au contacts and top gates (yellow), Si/SiO$_2$ substrate (light grey). **(b)** Optical micrograph showing one of our van der Waals heterostructures containing 6-layer InSe (dark yellow central region) which is sandwiched between thick (20-40nm) hBN crystals. The overlapping graphene contacts are outlined by the dashed green lines. Top gates serve to electrostatically define the quantum dot region and, also, tune the carrier density at the graphene/InSe interface. **(c,d)** Close-up SEM images of QD gates. Labels O, L, P, R stand for overall, left, plunger and right gates, respectively. Scale bars, 100 nm.

To define the QPCs, a negative voltage was applied to the top gates marked overall (O) and left (L) in Fig. 1d, meanwhile the other top gates (P and R) were held at zero potential. The total charge density was controlled by the back gate. Fig. 2a plots an example of the differential conductance G of such QPCs as a function of the split gate voltage $V_{LO}$ simultaneously applied to both O and L top gates (see Fig. 1d) at several fixed back-gate voltages $V_{BG}$. All of the curves exhibit step-like features close to integer multiples of the conductance quantum $2e^2/h \approx 77.46$ µS, which suggests 1D spin-degenerate channels ($h$ is the Planck constant and $e$ the elementary charge). The shift of the conductance curves towards negative $V_{LO}$ at higher $V_{BG}$ is due to the increasing electron density induced by the back gate. The quality of quantization is similar to, if not better than, that observed in graphene-based QPCs in zero magnetic field[21–23]. It is of note, that the step-like features are more pronounced at higher $n$ (dark blue curve in Fig. 2a), which makes it unlikely that mesoscopic conductance fluctuations contribute to these steps. The decrease in quality of the QPC features at lower $n$ can be attributed to shortening of the mean free path $l$ from ~190 nm at $V_{BG}$ = 70 V ($n$ = 3.5x10$^{12}$ cm$^{-2}$) to ~105 nm at 60 V ($n$ = 2.5x10$^{12}$ cm$^{-2}$).

The latter value is comparable to the size of our constrictions (see below). For the above estimates, we have used a reference device made in the Hall bar geometry and exhibiting similar μ (see SI), and applied the Drude formula $l = h\sigma/\sqrt{2\pi n}e^2$, where σ is the conductivity. The conductance steps were reproducible for different constrictions and devices however, the quantization quality could notably differ (orange curve in Fig. 2a), presumably because of disorder within the QPC regions.

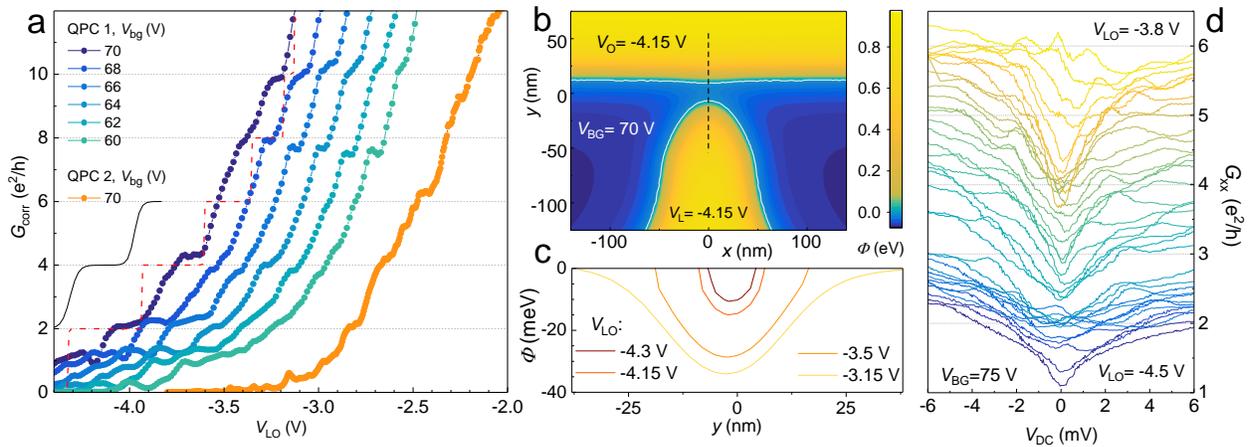

**Figure 2.** Conductance quantization in InSe point contacts. **(a)** Differential conductance of a QPC made from 6-layer InSe at 2K (lines with symbols). The series (parasitic) resistance $R_\mathrm{p} \approx 1$ kΩ was subtracted for each $V_{BG}$ (See Supplementary Information: Methods). Dashed red line: calculated positions and widths of plateaus in the ideal case. Solid black curve: plateau shapes expected in our devices. Orange curve: another QPC device. **(b)** Electrostatic simulation of a symmetric constriction, showing the offset ϕ of the conductance band. White lines: $\Phi$=0. **(c)** One dimensional potential extracted from (**b**) across the dashed line. **(d)** Uncorrected ($R_p$ = 0) conductance of QPC$_1$ at $V_{BG}$=75 V as a function of the applied DC voltage $V_{DC}$ along the constriction. $V_{LO}$ changes from -4.5 to -3.8 V in steps of 0.02 V (no offset). For all the measurements, the AC excitation was 100 μV, and 6 V was applied to the overlap gates above the InSe/graphene contacts.

To support our observations, we calculated numerically the 1D quantization conductance expected in our experimental constrictions, with SEM and AFM imaging being used to extract the size and shape of the constrictions. The self-consistent potential, $\Phi$, was calculated using the 3D Poisson equation with boundary conditions based on the electron density of states in the InSe layer:

$$\rho = -\frac{em}{\pi\hbar^2} e\Phi\, \mathrm{H}(\Phi),$$

Where $m = 0.14 m_e$ is the effective mass in 2D InSe[28], $m_e$ is the free electron mass, $\mathrm{H}(\Phi)$ is a step function that is zero when the potential is negative (fully depleted) and unity otherwise (when charge carriers are present) and $\Phi$ is essentially the Fermi energy with respect to the conduction band edge. The potential calculated for the QPC narrowing (dashed line in Fig. 2b) was then used to solve the 1D Schrödinger equation and find the number of transverse modes propagating through the QPC. Except for the QPC pinch-off voltage, this model has no adjustable parameters. The resulting conductance plateaus are shown by the red dashed curve in Fig. 2a.

For a more realistic description of the QPC, we used the parabolic saddle-point potential model[37]. The potential cuts were taken from our simulations of $\Phi$ for $V_{BG}$ = 70 V and then approximated with parabolas within the vicinity of the constriction center (Fig. 2b). This was repeated for different values of $V_{LO}$, which allowed us to find the dependence $G(V_{LO})$. A result of this modelling is shown in Fig. 2a by the black solid curve (for clarity, it is shifted by –0.5 V along the x-axis), showing good agreement with the experimental dependence. Note that our approximation of the parabolic saddle-point potential gets progressively worse with increasing $V_{LO}$ range.

For completeness, Fig. 2d shows the bias voltage spectroscopy for one of our QPCs at $V_{BG}$ =75 V and various split-gate voltages $V_{LO}$. These measurements allow us to estimate the energy spacing between the 1D subbands i.e. ⍰ ~ 5-10 meV, in good agreement with our solution to the 1D Schrödinger equation, which gives a separation of ⍰ =10 meV between the first and second 1D subband.

In further experiments, quantum dots were formed by the depletion of the 2D electron gas underneath all four top gates in Figs 1c,d. The coupling of such QDs to the source and drain electrodes was adjusted using the left and right barrier gates whereas the back gate voltage was fixed throughout the measurements. The plunger gate was used to tune the chemical potential inside the QDs. Initial tuning and symmetrization were carried out by recording the two-terminal differential conductance through the QD as a function of both $V_L$ and $V_R$, as shown in Fig. 3a. The conductance plot exhibits a series of bright lines that correspond to a finite conductance in the Coulomb blockade (CB) regime. We use this plot to select the gate voltages that lead to lithographically defined QDs and, thus, to avoid unintentional QDs due to disorder. The latter can appear inside the two constrictions. We expect that CB oscillations with a slope close to -1 in Fig. 3a originate from the intentional, lithographically defined QDs because the CB oscillations are equally affected by both gates. This indicates that the dot is localized somewhere between the two gates (an example is indicated by the white line in Fig. 3a). In contrast, horizontal or vertical lines in Fig. 3a correspond to unintentional QDs which are controlled by only one of the gates and, therefore, are localized in proximity. The latter QDs were discarded from our analysis. The region of gate voltages used in our analysis is marked by the circle in Fig. 3a.

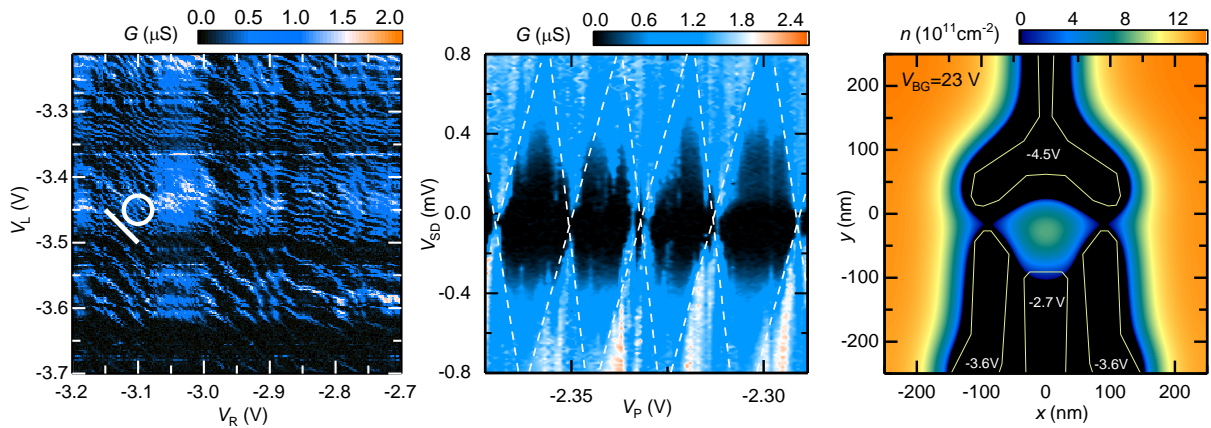

**Figure 3.** Gate-defined quantum dots in 2D InSe. **(a)** Map of the two-terminal differential conductance as a function of voltages $V_R$ and $V_L$ which control the left and right barriers. Other gates were fixed at $V_{BG}$ = 50 V, $V_O$= -4.51 V, $V_P$= -2 V; 10 µV AC excitation. The diagonal lines are due to gate defined QDs and therefore react to both side gates, as designed. Accidental QDs are caused by disorder and are more sensitive to one gate. **(b)** Conductance in the region highlighted by the circle in (a). Coulomb diamonds extend to the edges of the scanned range (dashed white lines). Temperature, 50 mK. **(c)** Charge carrier density found in our 3D electrostatic simulation. The gates are indicated by yellow lines.

Fig. 3b shows Coulomb diamonds observed in this region. The InSe QD exhibits an average plunger gate periodicity $\Delta V_P$ of ~19 mV. Using these measurements, we estimate an average charging energy $E_c$ of ~0.8 meV, which yields the dot's self-capacitance of $C_\Sigma = e^2/E_c$ of ~177 aF and plunger capacitance $C_P = e/\Delta V_P \approx 8$ aF. However, the irregular shape of the diamonds suggests that the charging energy is comparable to the QD's confinement energy $\Delta E$, which makes the above values only an approximation[38]. This is the case because the estimates are based on the assumption $\Delta E \ll E_c$. No additional resonance lines indicating excited states were observed within the conducting regions. This can be explained by a large tunneling amplitude $t$ over the left and right barriers at nonzero $V_{SD}$, which leads to a decrease in the lifetime of localized states and blurs fine features. The finite $t$ is also the reason why the conductance is finite within the diamonds ($t \ll E_c$ implies no conductance). We have also calculated a charge density map for realistic values of our gate voltages (Fig. 3c). The figure shows an island in the electron gas which has a size of ~100 nm and contains ~ 60 electrons, in qualitative agreement with the above estimate for $R$. Changing the plunger gate values in our model, we have found similar gate coupling of $C_P \approx 6$ aF

The found value of $C_\Sigma$ can be used to estimate the QD size by evoking the self-capacitance of an isolated disk with radius $R$ in a dielectric medium[11], $C_{disc} = 8R\varepsilon_0\varepsilon_r$, where $\varepsilon_0$ and $\varepsilon_r \approx 10$ are the permittivity's of a vacuum and InSe, respectively[39]. The total capacitance defining the charging energy is the sum of all the gate capacitances and the QD self-capacitance. For the purpose of our estimation, we assume that the barrier and plunger gate capacitances are approximately the same (~10 aF) and the overall gate capacitance is three times larger. Consequently, the self-capacitance of the InSe disk, $C_{disk}$, should be around 120 aF, yielding $R \approx 170$ nm.

In conclusion, we have demonstrated both one-dimensional and zero-dimensional confinement in few-layer InSe. The relatively small in-plane electron mass and high electron mobility in 2D InSe enables the one-dimensional quantization of electrons by electrostatic gating. In addition, using local top gates strategically placed over the InSe/graphene overlap, we have been able to significantly reduce contact resistance to InSe. Following from these promising results, it seems that two-dimensional materials are an excellent platform to study 1D and 0D physics. As such we envisage that InSe will play a strong role in the future of two-dimensional research and future applications.

**Corresponding Author**

*Email: Roman@Manchester.ac.uk, Matthew.Hamer@postgrad.Manchester.ac.uk

**Funding Sources**

This project has received funding from the European Union's Horizon 2020 research and innovation programme under grant agreement No 696656, and under the Marie Skłodowska-Curie grant agreement No 751883 (PTMCnano). In addition, funding has been received from the Engineering and Physical Sciences Research Council [grant number EP/M012700/1] (EPSRC).

**Acknowledgements**

We acknowledge the support of "The Wolfson Foundation", the "J P Moulton Charitable Foundation", and the "Garfield Weston Foundation" in establishing the IsoLab facility. We also acknowledge the National Academy of Sciences of Ukraine for providing the InSe.